\documentclass[%
 reprint,
 amsmath,amssymb,
 aps,
]{revtex4-2}

\usepackage{xcolor}
\usepackage{gensymb}
\usepackage{graphicx}
\usepackage{dcolumn}
\usepackage{bm}

\begin{document}

\preprint{APS/123-QED}

\title{Real time 3D coherent X-ray diffraction imaging}

\author{Fangzhou Ai}
\author{Vitaliy Lomakin}%
 \email{vlomakin@ucsd.edu}
\affiliation{%
Department of Electrical and Computer Engineering and Center for Memory and Recording Research, University of California, San Diego, La Jolla, CA 92093, U.S.A
}%

\author{Oleg Shpyrko}
\affiliation{
 Department of Physics, University of California, San Diego, La Jolla, CA 92093, U.S.A
}%

\begin{abstract}
Coherent X-ray Diffraction Imaging (CXDI) technique offers unique insights into the nanoscale world, enabling the reconstruction of 3D structures with a nanoscale resolution achieved through computational phase reconstruction from measured scattered intensity maps. Computational demands of 3D CXDI, however, limit its real-time application in experimental settings. This work presents a carousel phase retrieval algorithm (CPRA) that enables the real-time, high-resolution reconstruction of computationally complex 3D objects. CPRA is based on representing the 3D reconstruction problem as a set of 2D reconstructions of projected images corresponding to different experimentally collected angles via the Fourier slice theorem. Consistency between the 2D reconstructed images is based on an iterative procedure, in which each 2D reconstruction accounts for the adjacent 2D reconstructed images in a periodic (carousel) manner. Demonstrations on complex systems, including a lithium-rich layered oxide particle and a Staphylococcus aureus biological cell, demonstrate that CPRA significantly enhances the reconstruction quality and enables the reconstruction process to be completed in real time during experiment.

\end{abstract}


\maketitle

\begin{figure}
\includegraphics[scale=0.18]{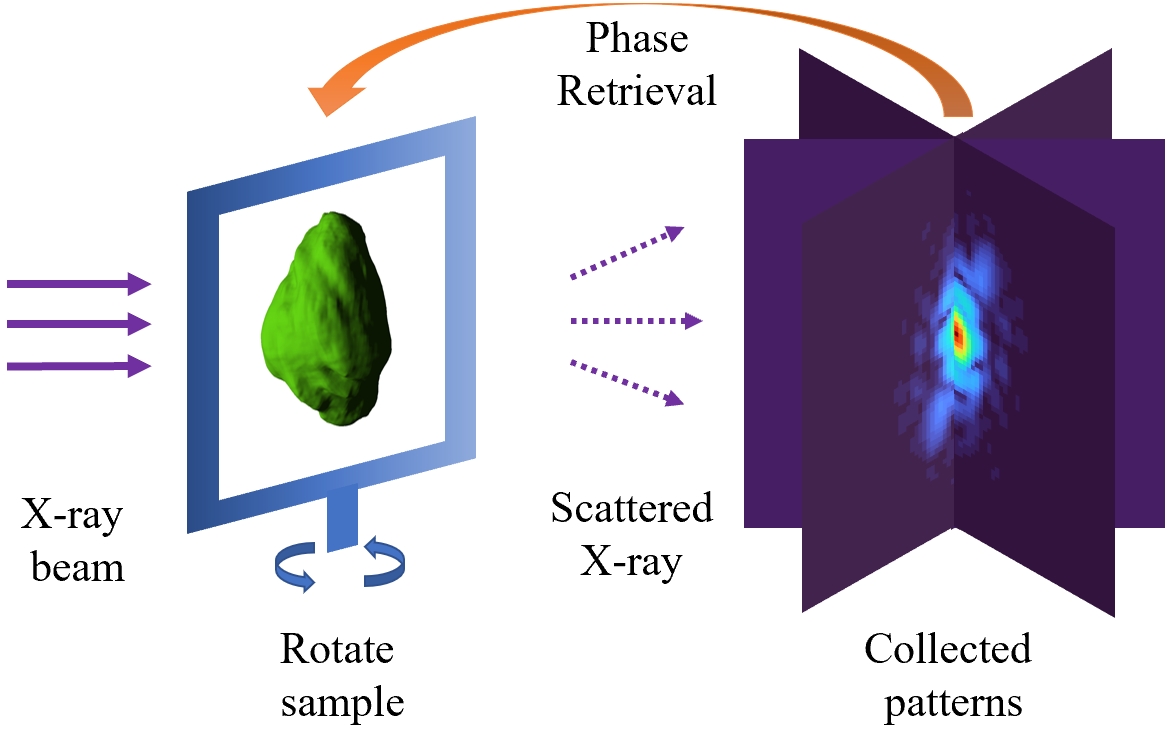}
\caption{\label{fig:1}Schematics of a 3D CXDI experimental and reconstruction procedure.}
\end{figure}

Coherent X-ray Diffraction Imaging (CXDI) is used to reconstruct 2D and 3D objects at a nanoscale resolution \cite{1}. 3D CXDI offers a non-invasive way to identify the object’s interior \cite{2} and finds many applications in physics, chemistry, and biology \cite{3,4,5,6,7,8,9,10,11,12,13,14,15,16}. With refined experiment techniques \cite{17,18} and algorithms \cite{19,20} CXDI is expected to play an increasingly important role across different disciplines. During an experiment, a coherent X-ray beam is diffracted by the object and a charge-coupled device (CCD) sensor collects the diffraction patterns as 2D images. For 3D object imaging, the object is rotated by a set of angles and the corresponding set of 2D images is collected. This procedure can be treated as a Fourier transform (FT) of the object as Fraunhofer’s diffraction \cite{2} (Fig.~\ref{fig:1}). The CCD sensor only collects the intensity whereas the phase information is lost. CXDI uses an iterative procedure based on forward and inverse FTs to extract the phase information for a high-resolution reconstruction \cite{21}.


Typically, 3D CXDI reconstruction is based on performing a 3D iterative phase retrieval algorithm \cite{20,21,22,23}, often with many iterations \cite{24} involving computationally intensive 3D forward and inverse FTs \cite{25,26,27}. The increasing experimental capabilities provide increasing collected data sizes, and highly efficient 3D CXDI algorithms become essential to assist experiments.  

We introduce a “carousel” phase retrieval algorithm (CPRA), which results in a major boost of the computational performance, accuracy, and robustness, thus allowing for a high-resolution reconstruction in real time during experiments. Instead of the direct 3D reconstruction, CPRA first retrieves the phases of the collected 2D diffraction patterns and then uses this information to reconstruct the 3D object. CPRA \cite{24,28} resolves shortcomings of inconsistencies of the 2D diffraction pattern reconstructions of related approaches \cite{24,28}, and thus allows for a rapid convergence.

In 3D CXDI, an object is rotated at a sequence of angles corresponding to incident wave vectors $\bm{q}_{i,n}$, where $n=1,...,N_p$ and $N_p$ is the total number of the angles (Fig.~\ref{fig:1}). For each $\bm{q}_{i,n}$, a scattered intensity map $I(\bm{q}_{t,n},\bm{q}_{i,n})$ of $N\times N$ pixels is collected. The map corresponds to a set of transverse wave vector components $\bm{q}_{t,n}=\bm{q}-\bm{q}_{i,n}$. Each $I(\bm{q}_{t,n},\bm{q}_{i,n})$ is related to a complex-valued spectral content $E(\bm{q}_{t,n},\bm{q}_{i,n})$ via $I(\bm{q}_{t,n},\bm{q}_{i,n})=|E(\bm{q}_{t,n},\bm{q}_{i,n})|^2$. A high reconstruction resolution requires $N_p$ and $N$ to be large. The goal is to reconstruct the object $O(\bm{r})$ in the real space, where $\bm{r}$ is the real space vector and $O(\bm{r})$ is the electron or spin density  \cite{14,29,30}. The phase information required for reconstruction needs to be recovered via a phase retrieval algorithm.

We now present CPRA with the aid of the Fourier slice theorem \cite{31}. We represent the measured 2D intensity for a given $\bm{q}_{i,n}$ as
\begin{eqnarray} \label{eq:1}
I(\bm{q}_{t,n},\bm{q}_{i,n})&&=\left|\hat{\bm{q}}_{i,n}\cdot \int \left[\int O(\bm{r})e^{i \bm{q} \cdot \bm{r}}d\bm{r}\right]d\bm{q}\right|^2,
\end{eqnarray}
where we have two integrals: an integral inside the square brackets is a 3D FT of the object $O(\bm{r})$ into its spectral content $E(\bm{q})$ and an outside integral over $d\bm{q}=\hat{\bm{x}}dq_x+\hat{\bm{y}}dq_y+\hat{\bm{z}}dq_z$ represents the object projected along the unit vector $\hat{\bm{q}}_{i,n}$ of $\bm{q}_{i,n}$. We, then, use the Fourier slice theorem \cite{31}, which, in this case, allows writing Eq.~\eqref{eq:1} as
\begin{eqnarray} \label{eq:2}
I(\bm{q}_{t,n},\bm{q}_{i,n}) = \left|\int \left[\hat{\bm{q}}_{i,n}\cdot \int O(\bm{r})d\bm{r}\right]e^{i\bm{q}_{t,n}\cdot \bm{r}_{t,n}}ds_{t,n}\right|^2
\end{eqnarray}
where $\bm{r}_{t,n}=\bm{r}-(\hat{\bm{q}}_{i,n}\cdot \bm{r})\hat{\bm{q}}_{i,n}$ is the coordinate and $ds_{t,n}$ is the surface differential in the plane transverse to the $\hat{\bm{q}}_{i,n}$ direction. Eq.~\eqref{eq:2} now has a 3D to 2D real-space projection inside the brackets, which is followed by a 2D FT. Let us regard the result in the square bracket as a new 2D projected object (PO) $O'(\bm{r}_{t,n}, \bm{q}_{i,n})=\hat{\bm{q}}_{i,n}\cdot \int O(\bm{r})d\bm{r}$ and formulate Eq.~\eqref{eq:2} in a 2D form as
\begin{eqnarray} \label{eq:3}
I(\bm{q}_{t,n},\bm{q}_{i,n}) = \left|\int O'(\bm{r}_{t,n}, \bm{q}_{i,n}) e^{i\bm{q}_{t,n}\cdot \bm{r}_{t,n}}ds_{t,n}\right|^2.
\end{eqnarray}
It follows that we can reconstruct a set of 2D POs $O'(\bm{r}_{t,n}, \bm{q}_{i,n})$ representing the projection of the object at the $n^{th}$ angle along $\hat{\bm{q}}_{i,n}$, which correspond to the discrete set of measured incident directions of $\bm{q}_{i,n}$ for the measured angles $\theta_{i,n}$. The reconstruction of each PO can be accomplished via the conventional CXDI procedure for the 2D case \cite{20} (also see Supplementary Materials).

Once $O'(\bm{r}_{t,n}, \bm{q}_{i,n})$ are obtained, the corresponding spectral contents $E(\bm{q}_{t,n},\bm{q}_{i,n})$ can be used for reconstructing the 3D object via tomography \cite{33}. The tomography procedure is based on $E(\bm{q})$, including its already recovered phase, which results in a fast convergence with a small number of initial random guesses.

A major difficulty of using CPRA is that the 2D reconstructed POs for different angles can be inconsistent \cite{24,28}. The POs are obtained from the same object, and they are required to point to the same object, while this is not guaranteed, e.g., they may have slightly different positions. These displacements may introduce errors when the 2D POs are combined for 3D reconstruction. Even there were a feasible alignment method \cite{34}, the reconstruction error or noise may, often unavoidably, have a detrimental effect on the reconstruction quality. Another, even more severe, factor is within the iterative phase retrieval algorithm itself. The algorithm requires many different random initial guesses. Different POs may converge to different local optima, and they cannot be combined for the final 3D reconstruction. Maintaining the 2D PO consistency is a key for enabling CPRA.

To resolve the 2D POs consistency challenge, we use the fact that the nearby 2D POs are similar as they correspond to close angles. We can maintain the reconstruction PO consistency and quality of an $n^{th}$ PO by using the $n-1^{th}$, $n+1^{th}$, $n-2^{th}$, $n+2^{th}$, etc., POs. This idea leads to CPRA. 

\begin{figure}
\includegraphics[scale=0.13]{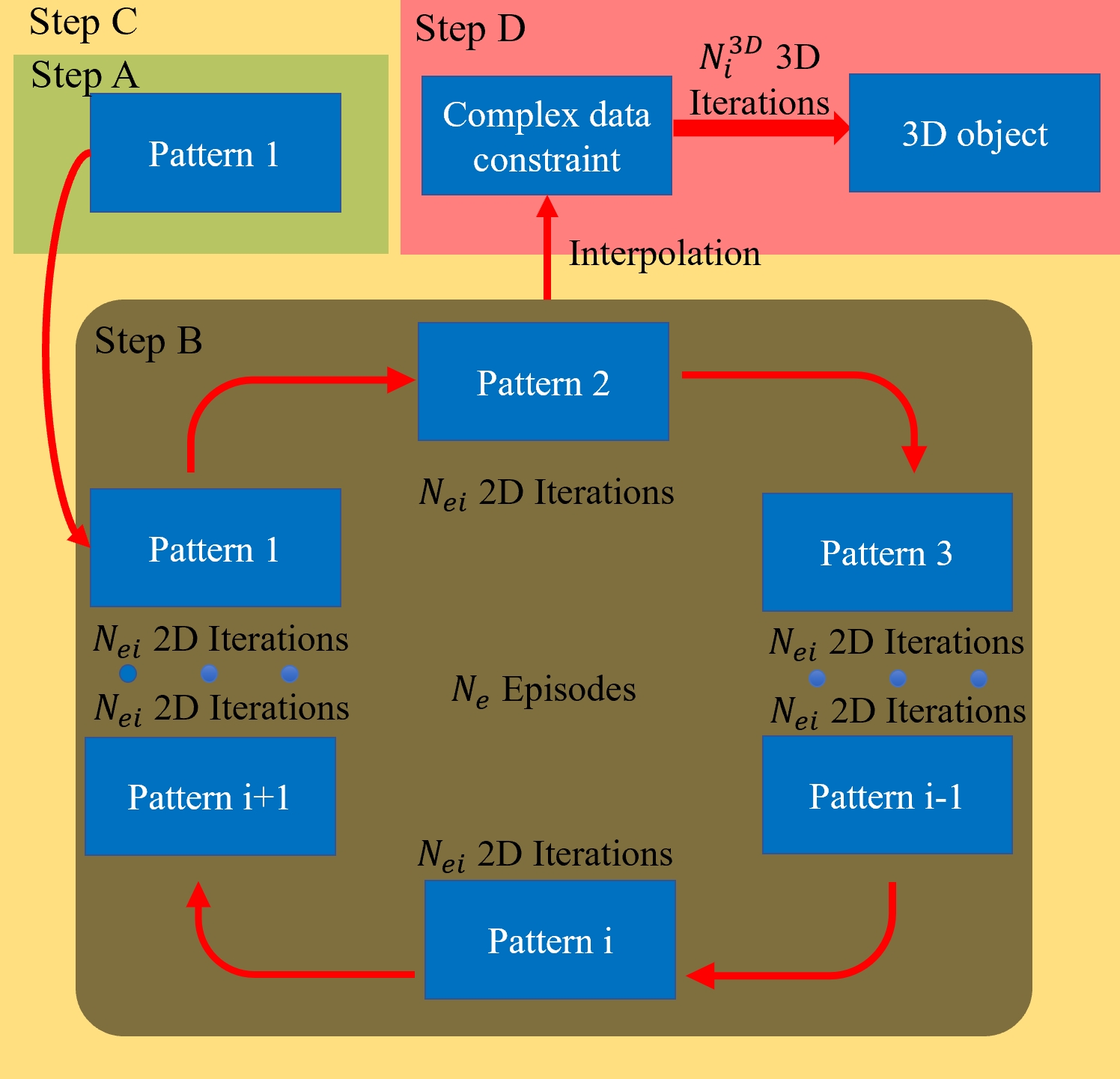}
\caption{\label{fig:2}Flowchart of the CPRA procedures.}
\end{figure}

Let us define the reconstructed POs as $O'_n=O'(\bm{r}_{t,n}, \bm{q}_{i,n})$ corresponding to $I_n=I(\bm{r}_{t,n}, \bm{q}_{i,n})$ and $E_n=E(\bm{r}_{t,n}, \bm{q}_{i,n})$. Let us call a set of POs for all $n=1,2,...,N_p$ as an episode. The $n^{th}$ intensities and spectral contents of an episode are periodic, such that $n=...,N_p-1,N_p,1,2,...$ correspond to consecutive incident angles. CPRA proceeds with an iterative process, in which an episode $W_{m+1}$ is obtained based on the previous episode $W_m$ via the following four steps, illustrated in Fig.~\ref{fig:2}:
\paragraph*{Step A: Pre-reconstruction} \label{step:1}
Initialize an episode $W_0$. Pre-reconstruct a single PO, e.g., for $n=1$, starting with a random initial guess. The pre-reconstruction is accomplished via the conventional 2D CXDI algorithm. This pre-reconstruction requires a relatively large number of iterations ($N_i^{pre}$), as the conventional 2D CXDI. The rest of the POs are initialized randomly. This initialization results in a set of POs $O_n^{\prime (0)}$ for the first episode $W_0$. 

\paragraph*{Step B: Iterative reconstructions of episodes} \label{step:2}
Reconstruct a set of episodes $W_m$ for $m=1,2,...,N_e$, where $N_e$ is the number of episodes. For each episode $W_m$, the POs are reconstructed via the conventional 2D CXDI consecutively, from $n=1$ to $n=N_p$, with the initial guess for each PO obtained based on the combination of the objects from episode $W_{m-1}$:
\begin{equation} \label{eq:4}
O^{\prime (m)}_n=\sum_{k=-N_p/2-n+1}^{N_p/2-n}\alpha_k O^{\prime (m-1)}_{k+n+N_p/2},
\end{equation}
where $\alpha_k$ represents weight coefficients that are periodic with respect to the range of $N_p$, satisfy $\sum_{k=-N_p/2+1}^{N_p/2}\alpha_k=1$, and are chosen such that they decrease with an increase of $|k|$. An example choice can be $\alpha_0=0.6$, $\alpha_{-1}=\alpha_1=0.2$ and $\alpha_k=0$ for $|k|>1$, otherwise. The fact that the coefficients are periodic implies that each reconstructed PO at a certain episode iteration is obtained based on adjacent objects from the previous episode iteration in a carousel (circular) manner. The reconstruction of the POs in all the episodes requires a small number of iterations $N_{ei}$ since this reconstruction is based on an approximation from the previous iteration, which is much better than a random guess. 
    
\paragraph*{Step C: Repeating and merging} \label{step:3}
Repeat step A and B multiple times, each for different random guesses in step A for the total number of random guesses $N_r$. We select a certain number $N_l<N_r$ of the best reconstructed episodes, which have the lowest reconstruction error. The POs with the same number may be inconsistent between episodes with different random guesses, caused by two factors. The first factor is because the projections of an object rotated along two parallel axes and displaced in the projected plane are the same. The displacement in the projected plane is proportional to the distance between the two axes. To address this issue, we move the mass center of each PO by $\Delta \bm{r}_n=-\int O'_n(\bm{r})\bm{r}dv$. The second factor comes from the property of FT that the intensity of an object’s FT is the same as the centrally inverted object, i.e., $|\int O'_n(\bm{r})e^{i\bm{q}\cdot \bm{r}}d\bm{r}|=|\int O'_n(-\bm{r})e^{i\bm{q}\cdot \bm{r}}d\bm{r}|$. To address this issue, we use the episode with the lowest reconstruction error. We calculate the Euclidean distance between one of the POs in this best episode and the same number PO in other episodes. We also calculate the Euclidean distance for the centrally inverted version of the POs in the episodes. We select the version with the lowest Euclidean distance, thus accomplishing the task of selecting the proper central symmetry. After obtaining the properly corrected episodes, we average between them to create the final episode, which is used as a set of $E_n$ for the final 3D reconstruction. 

\paragraph*{Step D: Final 3D object reconstruction} \label{step:4}
Interpolation and the final 3D reconstruction. We interpolate the final reconstructed $E_n$ to create a 3D spectral content $E(\bm{q})$ required for 3D reconstruction. We, proceed with a tomographic approach, which is like conventional 3D CXDI but using $E(\bm{q})$ obtained via the steps A-C of CPRA. The phase information in $E(\bm{q})$ leads to a rapid convergence and small number of required initial random guesses. Typically, only a single random guess and $N_i^{3D}$ (around 50) iterations are sufficient \cite{33}. 


The computational time of CPRA is dominated by step C, and it scales as $O(N_r(N_i^{pre}+N_pN_i^{2D})N^2\log N)$. The computational time of the conventional approach is due to 3D FTs at each iteration \cite{20} and it scales as $O(N_rN_i^{con}N^3\log N)$, where $N_i^{conv}$ is the number of 3D iterations that may be on the order of $10^3$ or even greater \cite{23}. Therefore, CPRA can be much more efficient by having $N_i^{pre}+N_pN_i^{2D}\ll N_i^{conv}N$ especially when the object size $N$ is large. In steps A-C, the memory consumption scales only as $O(N^2)$ as compared to the conventional 3D CXDI with a much greater $O(N^3)$ scaling. While step D requires $O(N^3)$ memory, it requires a small computational time scaled as $O(N_i^{3D}N^3\log N)$ as compared to the rest of CPRA. The memory consumption of step D may be reduced by performing 3D FTs as a set of 2D FTs, and the corresponding possible increase of the computational time associated with the need for the memory transfers is still insignificant. It shows that we can greatly accelerate CPRA by utilizing GPUs or additional parallelization even if $N$ is very large.

The following results show using CPRA for reconstructing a lithium-rich layered oxide particle based on computer-generated data as well as a Staphylococcus aureus (SA) cell based on experimentally collected data. The results are shown for CPU and GPU implementations of CPRA.

We start by considering an object that is uniformly sampled by $N\times N \times N$ pixels in the object domain. This object contains a lithium-rich layered oxide particle (the object in Fig.~\ref{fig:1}) \cite{35} that occupies less than $N/2$ pixels in each dimension to meet the oversampling requirements. The intensities corresponding to what is typically obtained from a synchrotron experiment are simulated by projections and FTs. The number of rotation angles $N_p$ is chosen to scale with $N$, such that $N_p/N=9/16$. We use these intensities to reconstruct the object via CPRA and conventional Shrinkwrap algorithm (CSWA) \cite{36}.

We first compare CPRA and CSWA when reconstructing the 3D object for a relatively small $N$, such that CSWA can be used. We, then, present the performance of the CPRA for large $N$, which is impossible to accomplish with CSWA. For the result verification, we choose $N=160$, $N_p=90$. We use the 2D CSWA method in steps A and B of CPRA, which makes the iterative algorithm identical for CPRA and CSWA. We remove pixels whose values are below a 0.1 relative defogging threshold \cite{33}. For assessing the reconstruction quality, we employ the Phase Retrieval Transfer Function (PRTF) \cite{23}, which evaluates the stability for different initial random guesses, and the Fourier Shell Correlation (FSC) \cite{37} between the original and reconstructed objects, which can be done because we have the ground truth original object. From both PRTF and FSC, we can estimate the reconstruction resolution, denoted as $R^{-1}$, by taking the cutoff spatial frequency at a threshold value, chosen as 0.5 \cite{23,37}.


\begin{figure}
\includegraphics[scale=0.12]{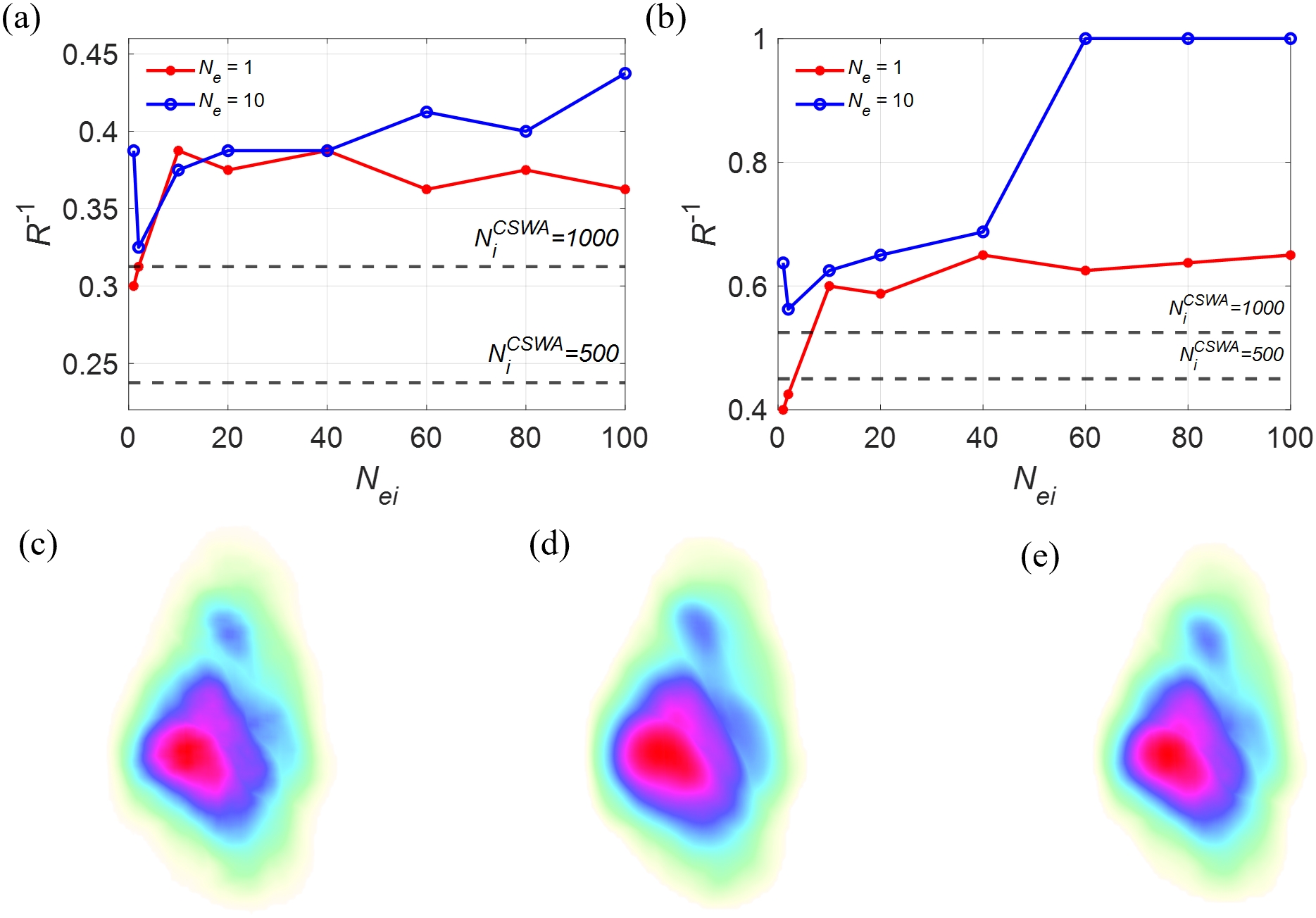}
\caption{\label{fig:3}Resolution determined by (a) PRTF and (b) FSC of CSWA and CPRA. (c) Corresponding volume density of the original object, (d) object reconstructed via CSWA (lower quality), and (e) object reconstructed via CPRA with $N_e=1$, $N_{ei}=10$ (higher quality).}
\end{figure}

Fig.~\ref{fig:3}(a) and~\ref{fig:3}(b) show PRTF and FSC as a function of the number of iterations in CPRA. These CPRA results are compared to those from CSWA for 500 and 1000 iterations. Similar to $N_i^{conv}$, this number can be defined as $N_i^{CSWA}$. We find that for CSWA with $N_i^{CSWA}=1000$, we can have more reasonable PRTF and FSC, which is consistent with other publication \cite{23}. CPRA, however, requires a much smaller number of iterations ($N_i^{2D}$) and has a much better performance. For instance, CPRA achieves the same resolution even when $N_e=1$, $N_{ei}=2$, and it achieves a much better resolution with $N_e=1$, $N_{ei}=10$.


To further compare the reconstruction qualities, we show 3D results (Fig.~\ref{fig:3}) for the original and reconstructed objects obtained via the CSWA and CPRA for $N_e=1$, $N_{ei}=2$ and $N_e=1$, $N_{ei}=10$. It is evident that CSWA is significantly less accurate, even visually, and that CPRA with $N_e=1$, $N_{ei}=10$ provides a better quality.


\begin{figure}
\includegraphics[scale=0.18]{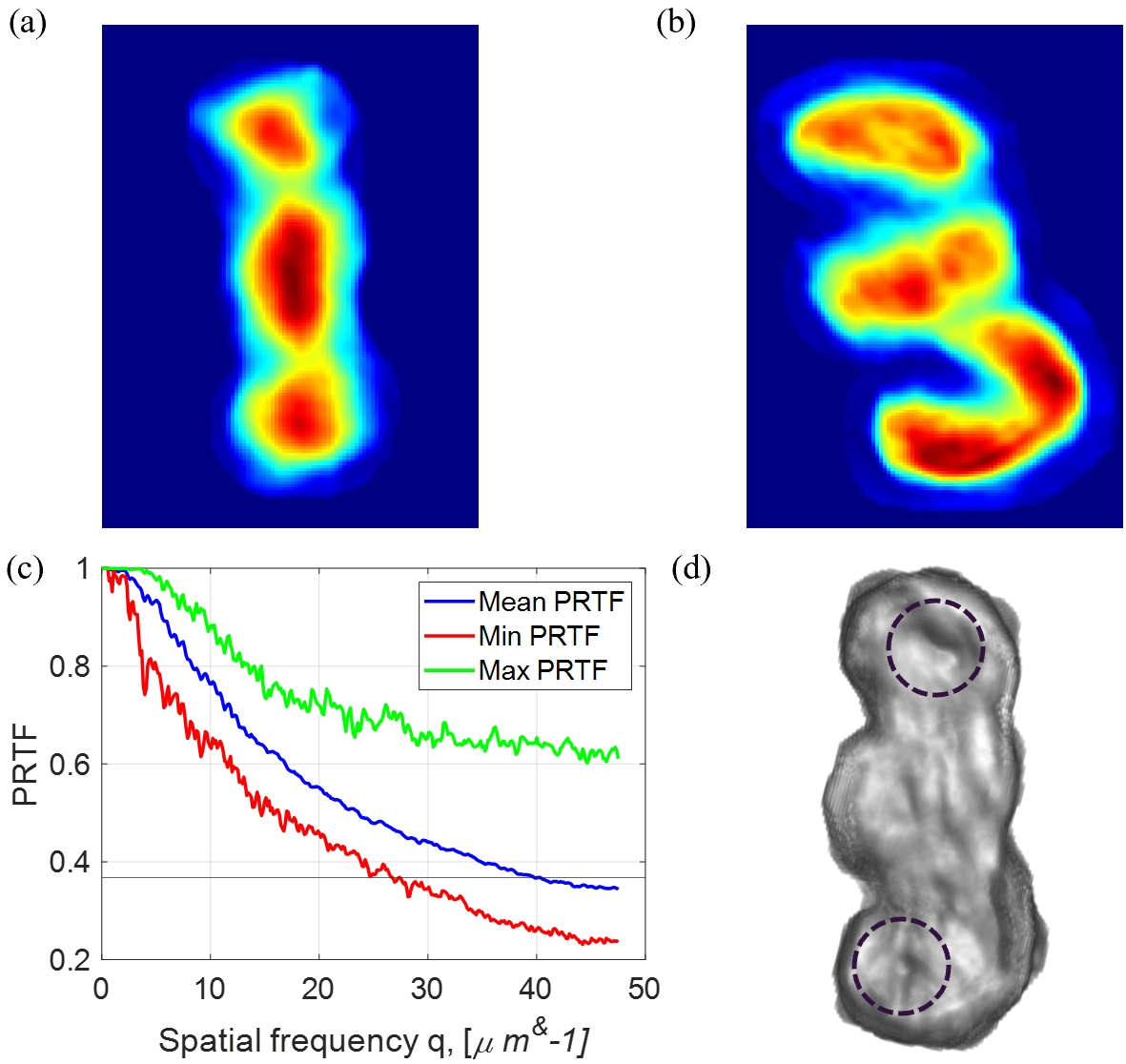}
\caption{\label{fig:5}Reconstruction of S. aureus cell: 2D reconstructed projected objects (a) $69.444\degree$ and (b) $0\degree$; (c) PRTF mean, minimal, and maximal values for all the 2D reconstructed objects (the horizontal line is the 1/e criterion); (d) 3D reconstructed surface morphology (black circles indicate the two representative depressions).}
\end{figure}

We, then, use CPRA with experimental data imaging Staphylococcus aureus (SA) cell \cite{38}. Each 2D intensity map is cropped to size $N \times N$, where $N=700$, with a beamstop in the center, and there are $N_p=27$ maps. The space constraints are pre-calculated by the Hybrid input-output (HIO) algorithm \cite{22} with the oversampling ratio of 4. To be consistent with the original reconstruction method \cite{38}, we adopt the Relaxed Averaged Alternating Reflections (RAAR) algorithm \cite{39} with hyperparameter $\beta=0.9$. For the pre-reconstruction step (step A), we perform $N_i^{RAAR}=500$ RAAR iterations on the first intensity map, half of the originally suggested number \cite{38}. We then set $N_e=2$, $N_{ei}=10$ for step B to reconstruct the rest of the 2D intensity maps. At the end of each episode, we perform an extra 2D CSWA iteration to obtain a tighter space constraint for each 2D image. We start from $N_r=100$ sets of random initial guesses and pick 10 best sets to merge. We project the 2D space constrains back to the 3D space to form a relatively tight 3D space constraint and perform 3D reconstruction via GENFIRE package \cite{33} with 50 iterations in total. We shrink the 3D space constraint every 5 iterations, leading to a tighter 3D space constraint close to the shape of the object at the end. We present the 2D reconstructions in Fig.~\ref{fig:5}(A-c), and the corresponding PRTF curve in Fig.~\ref{fig:5}(D) with $1/e$ criterion \cite{38}. In the Fig.~\ref{fig:5}(D), the mean value (blue line) of all 2D PRTFs suggests a very high resolution close to the achievable limit. The 3D reconstructed object surface morphology is presented in Fig.~\ref{fig:5}(E), where we find two representative depressions inside the black circles that agree with \cite{38}. 

Finally, we compare the computational performance of CPRA and CSWA on CPU and GPU computing architectures. We first consider the computational performance for reconstructing the object in Fig.\ref{fig:3}. The computations were performed on a desktop with a 16-core 4.9 GHz AMD R9-5950X CPU and NVIDIA RTX 3080Ti GPU. The shown CPU results were obtained on a single core. The multi-core CPU parallelization efficiency is above 90\%. The results are shown for $N=160,256,512,1024$ with corresponding $N_p=90,144,288,576$. For CSWA the number of iterations $N_i^{CSWA}$ was fixed at 1000, which is required for good reconstruction, and for CPRA, we set $N_e=1$, $N_{ei}=2$, which, according to the results in Fig.~\ref{fig:3}, gives a similar reconstruction quality. For both CSWA and CPRA, we set $N_r=100$. For large $N$, CSWA can take too much time, hence we run CSWA for only one random initial guess and then multiply this result by $N_r$ for performance comparisons.


\begin{figure}
\includegraphics[scale=0.13]{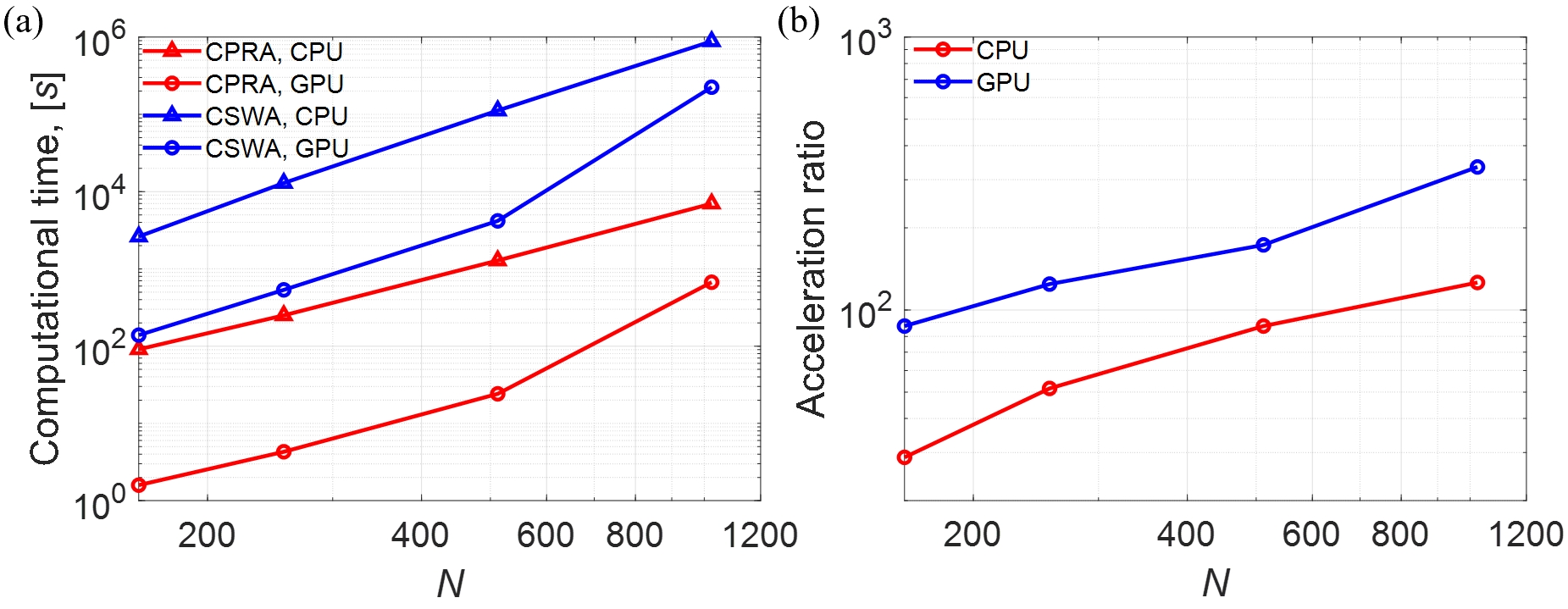}
\caption{\label{fig:6}Performance of CSWA and CPRA method. (a) computational time of different methods on CPU and GPU; (b) CPRA to CSWA acceleration ratio.}
\end{figure}

Fig.~\ref{fig:6}(a) presents the computational time for CPU and GPU computations using CPRA and CSWA, and Fig.\ref{fig:6}(b) shows the acceleration ratio of time consumption using CPRA versus CSWA on CPU and GPU. It is evident that CPRA is much faster for any $N$. The CPRA speed-up is in the range of 100-300 times on GPU and 30-120 times on CPU. The acceleration ratio increases with the problem size due better hardware utilization. As an example, reconstructing a case with $N=1024$ takes 670 sec, which allows CPRA to be used for real-time reconstruction during experiment.

We also compared the computational performance of CPRA for reconstructing the SA cell of Fig.~\ref{fig:5}. This case only required 100 initial guesses and 39 iterations for each 2D intensity map on average. CPRA took less than 20 sec on Nvidia RTX 3080 Ti, which is consistent with a real time reconstruction. CPRA was around 300 times faster than the conventional method when comparing on the same computing (CPU or GPU) architecture.

In summary, a highly efficient CPRA for 3D CXDI was introduced. CPRA requires a small number, as little as 1-2, of iterations and a small amount of memory to achieve a high-quality reconstruction. The numerical comparisons demonstrate that CPRA achieves 100-300 fold speed-ups on GPU, and 30-120 fold speed-ups on CPU with equal or even significantly higher reconstructing qualities. These performance allows executing 3D CXDI in real time, concurrent with experimental procedures. 

CPRA has several benefits that can allow future extensions, such as concurrent partial reconstructions or 2D POs and highly efficient implementations on multi-core, multi-CPU, and multi-GPU computing systems to reconstruct larger objects. Given the benefits of CPRA, it can become a part of a more efficient experimental apparatus in synchrotron facilities. The real-time 3D CXDI reconstruction can be combined with using numerical, e.g., micromagnetic \cite{14,40,41}, simulators, to further assist performing experiments. 

We thank Prof. Jianwei Miao, University of California, Los Angeles as well as Prof. Huaidong Jiang and Mr. He Bo, ShanghaiTech University for providing the experimental data and assistance in reconstruction of SA cell \cite{38}. The biological cell, figures, performance data can be acquired by contacting the corresponding author. The CPRA can be implemented on CPUs and GPUs, and the implementations are released on GitHub (\url{https://github.com/UCSD-CEM/Carousel-Phase-Retrieval-Algorithm}) as a header-only library under Apache License 2.0.

\bibliography{apssamp}

\end{document}